\documentclass[12pt]{article}
\usepackage{paper2e}
\usepackage{mydefs2e}
\usepackage{epsf}

\newcommand{\FI}{Fayet--Iliopoulos\xspace}

\begin{document}
\begin{titlepage}
\preprint{UMD-PP-97-116\\
{hep-ph/9706268}}

\title{Effective Lagrangians\\\medskip
and Light Gravitino Phenomenology}

\author{Markus A.~Luty\footnote{mluty@physics.umd.edu}\\
Eduardo Ponton\footnote{eponton@wam.umd.edu}}

\address{Department of Physics\\
University of Maryland\\
College Park, Maryland 20742, USA}

\begin{abstract}
We construct the low-energy effective lagrangian for
a light gravitino coupled to the minimal \susc standard model
under the assumption that \susy breaking is communicated to
the observable sector dominantly through soft terms.
Our effective lagrangian is written in terms of the
spin-$\frac{1}{2}$ Goldstino (the longitudinal component of the
gravitino) transforming under a non-linear realization of \susy.
In this lagrangian, the Goldstino is derivatively coupled and
all couplings of the Goldstino to light fields are determined
uniquely by the \susy-breaking scale $\sqrt{F}$.
This lagrangian is therefore a useful starting point for
further investigation of the light gravitino in gauge-mediated
\susy breaking models.
We show that the invisible width of the $Z$ into Goldstinos gives
the constraint $\sqrt{F} \gsim 140\GeV$.
\end{abstract}

\date{Revised version December 1997}

\end{titlepage}

% ----------------------------------------------------------------------------
% Start of main Text
% ----------------------------------------------------------------------------
\section{Introduction}
If \susy plays a role in solving the gauge hierarchy problem, the 
scale $M_S$ of the masses of superpartners of observed particles
cannot be much larger than $1\TeV$.
However, the scale $\sqrt{F}$ at which \susy is spontaneously
broken can be much larger \cite{OldLow,LargeSUSY}.%
\footnote{We follow conventional practice and parameterize the scale of
\susy breaking by a quantity $F$ that has dimension mass squared.}
Clearly it is important and interesting to obtain experimental
information about the scale $\sqrt{F}$.

One of the few direct experimental handles on this scale comes from the
fact that spontaneous breaking of \susy gives rise to a massless
spin-$\frac{1}{2}$ Goldstino, the Nambu--Goldstone mode associated
with \susy breaking.
The presence of supergravity means that the Goldstino is absorbed
as the longitudinal component of the spin-$\frac{3}{2}$ gravitino
(superpartner of the graviton), giving rise to a spin-$\frac{3}{2}$
particle with mass \cite{SuperHiggs}
\beq[gravmass]
m_{3/2} = (2.5 \times 10^{-4} \eV) \left( \frac{F}{(1 \TeV)^2} \right).
\eeq
In this paper, we consider models in which the gravitino mass is small
compared to $M_S$, so that the gravitino is the lightest \susc particle.
This includes both gauge-mediated models \cite{OldLow,NewLow}
and ``low-scale'' supergravity models \cite{noscale}.

We will assume that \susy breaking is communicated to the observable
sector via some ``messenger'' interactions, and that the dominant
source of \susy breaking in the observable sector comes through
soft terms.
Under this assumption, the couplings of the gravitino to matter are
completely determined in terms of the \susy breaking scale $\sqrt{F}$.
This is in contrast to a completely model-independent approach, in which
one writes the most general couplings of matter to a gravitino
constrained only by non-linearly realized \susy.
In such a lagrangian, one finds that some of the couplings of the
Goldstino are not determined by the Goldstino-independent amplitudes.
The results of this paper can be interpreted as giving the values for
these couplings in a large and interesting class of models.
Deviations from these values would presumably occur only if \susy was
broken directly in the observable sector, or if ``hard'' breaking of
\susy was important.
Our effective lagrangian can be used as the starting point for further
phenomenological study of the gravitino in the class of models we
consider.

We use our effective lagrangian to place bounds on $\sqrt{F}$.
This subject has a long history going back to the earliest days
of \susy \cite{Fayet}.
We find that the couplings of the gravitino are suppressed by
additional powers of $E/\sqrt{F}$ compared to couplings used
by some authors, invalidating some of the bounds
in the literature \cite{others,Rabi}.
An earlier version of this paper found an additional coupling of
a photon to gravitino pairs of order $M^2 E^2 / F^2$, where $M \sim 50\GeV$
is the coefficient of a \FI $D$ term;
we now find that this term is not present, in agreement with \Ref{Purd}.
We obtain a bound on $\sqrt{F}$ from the invisible
width of the $Z$ to gravitinos:
$\sqrt{F} \gsim 150\GeV$.
This is far from ruling out gauge-mediated or low-scale models.
However, we expect similar bounds (involving unknown order-1
coefficients) in any model.
Naturalness constraints tell us only that
$\sqrt{F} \gsim 100\GeV$, and we believe it is interesting that the
bounds obtained in this paper begin to explore this range.

This paper is organized as follows.
In Section 2, we derive the effective lagrangian using a non-linear
realization of \susy.
In Section 3, we derive bounds on $\sqrt{F}$ from $Z$ decay.
Section 4 contains our conclusions.

% ----------------------------------------------------------------------------
\section{Effective Lagrangian}
% ----------------------------------------------------------------------------
The interactions of the Goldstino at low energies are governed by 
low-energy theorems analogous to those which apply when global 
internal symmetries are spontaneously broken.
The most convenient formalism for analyzing these interactions is an 
effective theory in which the broken symmetry (in this case 
\susy) is realized non-linearly.
This approach ensures that our results apply to any model in which
\susy is broken spontaneously.
In this section, we review the non-linear realization of \susy of 
\Refs{Volkov,SamWess} and adapt it to the case of \susy breaking
in a hidden sector.
(For an alternate formalism, see \Ref{OtherGold}.)

\Susy can be non-linearly realized on a single spinor field $\hat{\chi}$
via%
\footnote{We use the spinor conventions of Wess and Bagger \cite{WB}.}
\beq[chiSUSYtrans]
(\xi Q + \bar{\xi} \bar{Q})
\times \hat{\chi}_{\al}(x) = \xi_{\al}
- i \left[ \hat{\chi}(x) \si^{\mu} \bar{\xi}
- \xi \si^{\mu} \bar{\hat{\chi}}(x)
\right] \partial_{\mu} \hat{\chi}_{\al}(x).
\eeq
With this definition, $\hat{\chi}$ has dimension $-\frac{1}{2}$,
the same as the superspace coordinates $\th$.

To construct \susc lagrangians involving $\hat{\chi}$, we follow
\Ref{SamWess}.
The field $\hat{\chi}$ can be promoted to a superfield $\Th$ via
\beq
\Th^{\al}(x, \th, \bar{\th}) \equiv e^{\th Q + \bar{\th} Q}
\times \hat{\chi}^{\al}(x)
= \th^{\al} + \hat{\chi}^{\al}(x) + \cdots,
\eeq
where the generators $Q$ act on $\hat{\chi}$ according to
\Eq{chiSUSYtrans}.
The resulting superfield satisfies
\beq[proj]
\bal
D_{\al} \Th^{\be} &= \de_{\al}{}^{\be} + i \bar{\Th}^{\dot{\al}}
\si^{\mu}_{\al \dot{\al}} \partial_{\mu} \Th^{\be},
\\
\bar{D}_{\dot{\al}} \Th^{\be} &= -i \Th^{\al} \si^{\mu}_{\al \dot{\al}}
\partial_{\mu} \Th^{\be}.
\eal
\eeq

We can use the superfield $\Th$ to write manifestly \susc
interactions for the field $\hat{\chi}$.
The most general effective lagrangian for $\hat{\chi}$ has the form
\beq
\scr{L} = \myint d^{2}\th d^{2}\bar{\th}\,
\scr{F}(\Th, \bar{\Th}, D_{\al}\Th, \ldots).
\eeq
The kinetic term for $\hat{\chi}$ is contained in the term%
\footnote{The term
$\int d^{2}\th d^{2}\bar{\th}\,
\left( \Th^{2} + \bar{\Th}^{2} \right)$
is a total spacetime derivative.}
\beq[VALag]
\scr{L}_{\rm AV} = -\frac{F^{2}}{2} \myint d^{2}\th d^{2}\bar{\th}\,
\Th^{2} \bar{\Th}^{2}
= -2 F^{2} \hat{\chi} \si^{\mu} \partial_{\mu} \bar{\hat{\chi}}
+ \cdots,
\eeq
which is the Volkov--Akulov lagrangian.
The constant $F$ has dimension 2, and parameterizes the scale of
\susy breaking.
The canonically normalized Goldstino field is given by
\beq[canonchi]
\chi \equiv \sqrt{2} F \hat{\chi}.
\eeq

The lagrangian also contains higher order terms such as
\beq[Higher]
\de\scr{L} = c F \myint d^{2}\th d^{2}\bar{\th}\,
D^{2} \bar{\Th}^{2} \bar{D}^{2} \Th^{2}
\sim \frac{c}{F^3}
\left| \partial_{\mu} \chi \si^{\mu\nu}
\partial_{\nu} \bar{\chi} \right|^{2}
+ \cdots,
\eeq
where $c$ is a dimensionless coupling constant.
This gives rise to a systematic low-energy expansion for
Goldstino interactions in powers of the Goldstino energy.
The parameter $c$ is somewhat model-dependent.
For example, if the dynamics that breaks \susy is strongly coupled
with no small parameters, the effective lagrangian has the form
\cite{fourpi}
\beq
\scr{L} \sim F^2 \sum_{p, \ldots, r} \,
\myint d^{2}\th d^{2}\bar{\th}\,
\left( \frac{D_\al\vphantom{\bar{D}_{\dot{\al}}}}{\La^{1/2}} \right)^p
\left( \frac{\bar{D}_{\dot{\al}}}{\La^{1/2}} \right)^{\bar{p}}
\left( \frac{\partial_\mu}{\La} \right)^q
\left( \La^{1/2} \bar{\Th} \right)^{\bar{r}}
\left( \La^{1/2} \Th \vphantom{\bar{\Th}}  \right)^r,
\eeq
where $\La \sim \sqrt{4\pi F}$ is the scale of the strong resonances.
In such a model, $c \sim 1/(4\pi)$.

We now consider the coupling of the Goldstino to matter in
models where $\sqrt{F} \gg M_S$ and \susy is communicated to the
observable fields by some ``messenger'' sector.
We assume that all fields in the messenger sector are heavy
compared to $M_S$, and so the only light degrees of freedom are
the fields of the \susc standard model and the Goldstino.
We further assume that the \susy breaking in the observable sector
occurs dominantly through soft \susy breaking terms.
This is the case in both gauge-mediated and gravity-mediated
\susy breaking (including ``no-scale'' models).
%
% models with $\sqrt{F} \gg M_{S}$, where $M_S$ is the mass scale
% of the superpartners of the observed particles.
% Our aim is to construct the effective lagrangian
% at the scale $M_{S}$ in such models.
% The light fields will be the observable sector fields and
% a Goldstino.
% The full theory may involve a hidden sector as well as
% ``messenger'' interactions that communicate the \susy breaking to the
% observable sector;
% we assume that all fields in this sector are heavy compared to $M_{S}$.
% We then expect that the largest \susy-breaking terms in the 
% observable sector are those that break \susy softly
%\cite{LargeSUSY}.
%
These terms can be written in terms of \susy-breaking ``spurion'' 
fields such as $\th\th\bar{\th}\bar{\th}$ \cite{SUSYSpurion}.
In order to make these terms \susc, we simply write the spurions in 
terms of the superfields $\Th$ in place of the superspace 
coordinates $\th$.
The most general soft breaking terms can then be written
as follows:%
\footnote{Some soft terms are conventionally written as
$d^{2}\th$ integrals.
The $\Th$ superfield is not chiral, so we must write these terms as full 
superspace integrals to make them \susc.}
\beq[SoftL]
\scr{L}\rsub{soft} = \myint d^{2}\th d^{2}\bar{\th}\,
\Th^{2} \bar{\Th}^{2}\, \scr{O}\rsub{soft},
\eeq
where
\beq[SoftTerms]
\bal
\scr{O}\rsub{soft} &=
-(m^{2})^{a}{}_{b}
(\bar{\Phi} e^{2 g V_A T_A} \Phi)_{a}{}^{b}
- \frac{1}{2} \left[
B_{ab} \Phi^{a} \Phi^{b} + \hc \right]
\\
&\quad + \frac{1}{2} 
\tilde{M}_A \left( W^{\al}_A W^{\vphantom{\al}}_{\al A} + \hc \right)
+ \frac{1}{6}
\left[ A_{abc} \Phi^{a} \Phi^{b} \Phi^{c} + \hc \right]
\\
&\quad + \frac{1}{4} \left[ g_1 M_D^2 D^\al W_{\al 1} + \hc \right].
\eal\eeq
The last term is an induced \FI term (see \Eq{DtermComp} below)
that can appear when there is a $\U1$ factor in the gauge group.
This term should be counted as a soft \susy breaking term, since it does
not lead to new quadratic divergences.
Although it is possible that symmetries forbid such a term, contributions
of this type are induced in general models after \susy breaking, as we
discuss below.%

We can find the component field expression of the terms above by 
projection with the help of \Eqs{proj}.
We are interested only in terms containing at most two Goldstino 
fields, and we obtain
\newcommand{\os}{\left.\scr{O}\rsub{soft}\right|_{0}}
\beq[SoftEffL]
\bal
\de\scr{L}\rsub{soft} &= \os
- \left( \hat{\chi}^{\al} D_{\al} \os + \hc
\right)
\\
&\quad - \frac{1}{4} \left( \hat{\chi}^{2} D^{2} \os + \hc
\right)
\\
&\quad + \frac{1}{2} \hat{\chi}^{\al} \bar{\hat{\chi}}^{\dot{\al}}
[ D_{\al}, \bar{D}_{\dot{\al}} ] \os + O(\hat{\chi}^{3}),
\eal\eeq
where $|_{0}$ denotes the $\th = \bar{\th} = 0$ component.
In writing this result, we have omitted terms
that vanish by the lowest-order $\hat{\chi}$ equations of motion.%
\footnote{More precisely, these terms can be eliminated by a field 
redefinition to the order we are working.}

It is straightforward to compute the component form of
\Eq{SoftEffL}.
The scalar masses of \Eq{SoftTerms} give
\beq[DiracSoft]
\bal
\de\scr{L}\rsub{soft} &= -(m^{2})^{a}{}_{b} \Bigl[
\bar{\phi}_{a} \phi^{b}
- \sqrt{2} \left( \bar{\phi}_{a} \hat{\chi} \psi^{b} +
\bar{\hat{\chi}} \bar{\psi}_{a} \phi^{b} \right)
\\
&\qquad\qquad\qquad
+ \left( \hat{\chi}^{2} \bar{\phi}_{a} F^{b} +
\bar{\hat{\chi}}^{2} \bar{F}_{a} \phi^{b} \right)
+ i \hat{\chi} \si^{\mu} \bar{\hat{\chi}}
(\bar{\phi} \scr{D}_{\mu} \phi -
\scr{D}_{\mu} \bar{\phi} \phi)_{a}{}^{b}
\\
&\qquad\qquad\qquad
+ 2 (\bar{\hat{\chi}} \bar{\psi}_{a}) (\hat{\chi} \psi^{b})
\Bigr] + O(\hat{\chi}^{3}).
\eal\eeq
The ``$B$ term'' scalar masses give
\beq[MajoranaSoft]
\bal
\de\scr{L}\rsub{soft} &= -\frac{1}{2} B_{ab}
\Bigl[ \phi^{a} \phi^{b}
- \sqrt{2} \left( \hat{\chi} \psi^{a} \phi^{b}
+ \hat{\chi} \phi^{a} \psi^{b} \right)
\\
&\qquad\qquad\qquad\quad
- \hat{\chi}^{2} \left( \psi^{a} \psi^{b} - F^{a} \phi^{b}
- \phi^{a} F^{b} \right)
\Bigr] + \hc + O(\hat{\chi}^{3}).
\eal\eeq
The gaugino mass terms give
\beq[GauginoSoft]
\bal
\de\scr{L}\rsub{soft} &= \frac{1}{2} \tilde{M}_A \Bigl[
-\la_A \la_A
+ 2 \hat{\chi} \si^{\mu\nu} \la_A F_{\mu\nu A}
+ 2i \hat{\chi} \la_A D_A
\\
&\qquad\qquad
- \frac{1}{2} \hat{\chi}^{2}
\left(F^{\mu\nu}_A F_{\mu\nu A} +
i F^{\mu\nu}_A \tilde{F}_{\mu\nu A}
- 2 D_A D_A + 4i \la_A \si^{\mu} (\scr{D}_{\mu} \bar{\la})_A
\right)\Bigr]\ \ \ \ \ 
\\
&\qquad\qquad
+ \hc + O(\hat{\chi}^{3}),
\eal\eeq
where $\tilde{F}^{\mu\nu} = \frac{1}{2} \ep^{\mu\nu\la\rho} F_{\la\rho}$.
The trilinear terms give
\beq[TrilinearSoft]
\bal
\de\scr{L}\rsub{soft} &= \frac{1}{6} A_{abc} \Bigl[
\phi^{a} \phi^{b} \phi^{c}
- \sqrt{2} \left( \hat{\chi} \psi^{a} \phi^{b} \phi^{c}
+ \phi^{a} \hat{\chi} \psi^{b} \phi^{c}
+ \phi^{a} \phi^{b} \hat{\chi} \psi^{c} \right)
\\
&\qquad\qquad\quad
- \hat{\chi}^{2} \bigl( \phi^{a} \psi^{b} \psi^{c}
+ \psi^{a} \phi^{b} \psi^{c} + \psi^{a} \psi^{b} \phi^{c}
\\
&\qquad\qquad\qquad\qquad
- F^{a} \phi^{b} \phi^{c} - \phi^{a} F^{b} \phi^{c}
- \phi^{a} \phi^{b} F^{c} \bigr)
\Bigr] + \hc + O(\hat{\chi}^{3}).
\eal\eeq
Finally, the induced \FI term gives
\beq[DtermComp]
\de\scr{L}\rsub{soft} = g M_D^2 \left[ D
- (\partial^\mu \hat{\chi} \si^\nu \bar{\hat{\chi}} + \hc) F_{\mu\nu}
\right] + O(\chi^3).
\eeq

In the lagrangian of \Eqs{SoftEffL} through
\eq{DtermComp}, the Goldstino field $\hat{\chi}$ is not
derivatively coupled.
This form of the lagrangian is useful in the energy regime
$M_{\rm S} \ll E \ll F$, where $M_{\rm S}$ is the scale of the
soft \susy breaking masses.
In this regime, \Eq{SoftEffL} shows that the couplings of the
Goldstino to observable matter fields are suppressed by positive
powers of $M_{\rm S}$.

If one is interested in the phenomenology of the
Goldstino in the energy regime $E \sim M_S$ (or $E \ll M_S$),
it is convenient to perform a field redefinition to obtain a lagrangian
in which the Goldstino field is derivatively coupled.
This makes manifest the fact that the couplings of the Goldstino are
suppressed by positive powers of $E$ for $E \ll M_{S}$.
We make this field redefinition by making a \emph{local} \susy
transformation on the matter fields
with parameter $-\hat{\chi}(x)$.
In Wess--Zumino gauge, the \susy transformations on component
fields are generated by the operators $Q_{\rm WZ}$ defined by
\cite{WB}
\beq
\bal
(\xi Q_{\rm WZ} + \bar{\xi} \bar{Q}_{\rm WZ}) \times \phi^a
&= \sqrt{2} \xi \psi^a,
\\
(\xi Q_{\rm WZ} + \bar{\xi} \bar{Q}_{\rm WZ}) \times \psi^a
&= i\sqrt{2} \si^\mu \bar{\xi} (\scr{D}_\mu \phi)^a
+ \sqrt{2} \xi F^a,
\\
(\xi Q_{\rm WZ} + \bar{\xi} \bar{Q}_{\rm WZ}) \times F^a
&= i\sqrt{2} \bar{\xi} \bar{\si}^\mu (\scr{D}_\mu \psi)^a
+ 2 i g \bar{\xi} (\bar{\la} \phi)^a,
\\
(\xi Q_{\rm WZ} + \bar{\xi} \bar{Q}_{\rm WZ}) \times A_{\mu A}
&= -i \bar{\la}_A \bar{\si}_\mu \xi
+ i \bar{\xi} \bar{\si}_\mu \la_A,
\\
(\xi Q_{\rm WZ} + \bar{\xi} \bar{Q}_{\rm WZ}) \times \la_A
&= \si^{\mu\nu} \xi F_{\mu\nu A} + i \xi D_A,
\\
(\xi Q_{\rm WZ} + \bar{\xi} \bar{Q}_{\rm WZ}) \times D_A
&= -\xi \si^\mu (\scr{D}_\mu \bar{\la})_A
- (\scr{D}_\mu \la)_A \si^\mu \bar{\xi}.
\eal
\eeq
We then define new fields by the field redefinition
\beq[fieldtrans]
\bal
\Phi^{a} &= \left.
e^{\xi Q_{\rm WZ} + \bar{\xi} \bar{Q}_{\rm WZ}}
\times \Phi^{\prime a} \right|_{\xi = \hat{\chi}},
\\
V_A &= \left.
e^{\xi Q_{\rm WZ} + \bar{\xi} \bar{Q}_{\rm WZ}}
\times V'_A \right|_{\xi = \hat{\chi}},
\\
\hat{\chi} &= \hat{\chi}'.
\eal
\eeq

When the lagrangian is written in terms of the primed
fields defined above, the Goldstino is derivatively coupled.
To establish this to all orders in $\hat{\chi}$,
we will show that when $\hat{\chi}$ is taken to be a
constant, the lagrangian is independent of $\hat{\chi}$.
The crucial observation is that the \trans between the primed and
unprimed field is ``almost'' a \susy \trans.
If $\hat{\chi}$ were a constant, and we combined the change of variables
above with a \trans \Eq{chiSUSYtrans} acting on $\hat{\chi}$ with parameter
$\xi = \hat{\chi}$, the result would be a (nonlinear) \susy \trans under which
the lagrangian is fully invariant.
Therefore, for constant $\hat{\chi}$, the
change of variables \Eq{fieldtrans} is equivalent to
\beq
\bal
\Phi^{a} &= \Phi^{\prime a}, \qquad
V_A = V'_A,
\\
\hat{\chi}
&= \left. e^{-(\xi Q + \bar{\xi} \bar{Q})} 
\times \hat{\chi}' \right|_{\xi = \hat{\chi}}
\equiv 0
\quad \hbox{(for $\hat{\chi} = \hbox{constant}$)}.
\eal
\eeq
The fact that the $\hat{\chi}$ fields are set to zero by this \trans
follows directly from \Eq{chiSUSYtrans}, and shows that the lagrangian
is independent of $\hat{\chi}$ for constant $\hat{\chi}$.

When we compute the lagrangian in terms of the fields defined in
\Eq{fieldtrans}, we find that there are still Goldstino couplings
proportional to the gaugino mass $M$.
It is convenient to eliminate these by a further field redefinition
\beq[morefieldtrans]
\la^{\prime}_A = \la_A^{\prime\prime} - \frac{i}{2} \left[
( \hat{\chi} \si^\mu \bar{\la}_A^{\prime\prime}) \partial_\mu \hat{\chi}
+ (\hat{\chi} \partial_\mu \hat{\chi}) (\bar{\la}_A^{\prime\prime} \bar{\si}^\mu )
\right].
\eeq
When the matter field kinetic terms are written in terms of the primed
fields defined in \Eqs{fieldtrans} and \eq{morefieldtrans}, we obtain
derivative couplings of the Goldstino to the matter fields with no
couplings proportional to soft \susy-breaking masses.
(The superpotential terms contain no derivatives, and therefore do not
induce Goldstino couplings.)
We will assume that these kinetic terms are canonical, as is appropriate
for gauge-mediated \susy breaking.%
\footnote{It is straightforward to generalize our results to an arbitrary
\Kahler potential.}
The final lagrangian can then be written
\beq
\scr{L}_{\rm eff} = \sum_n \frac{1}{F^n} \left[
\scr{L}^{(n)}_{\rm matter}
+ \scr{L}^{(n)}_{\rm gauge}
+\scr{L}_{\rm soft}^{(n)} \right],
\eeq
where $n$ counts the number of powers of $\hat{\chi}$.
Expressing the results in terms of the canonically normalized
field $\chi$ defined in \Eq{canonchi} and dropping the primes,
the matter coupling terms give the usual \susc terms
\beq[matterzero]
\bal
\scr{L}^{(0)}_{\rm matter} &=
-(\scr{D}^\mu \phi)^a (\scr{D}_\mu \bar{\phi})_a
- i \psi^a \si^\mu (\scr{D}_\mu \bar{\psi})_a
+ F^a \bar{F}_a
\\
&\qquad
+ i\sqrt{2} g_A [\psi^a (\la_A \bar{\phi})_a
- \phi^a (\bar{\la}_A \bar{\psi})_a]
- \frac{g_A^2}{2} (\bar{\phi} T_A \phi)^2
\\
&\qquad + \biggl[
\frac{\partial W(\phi)}{\partial\phi^a} F^a
+ \frac{\partial^2 W(\phi)}{\partial\phi^a \partial\phi^b}
\left( \phi^a F^b - \sfrac{1}{2} \psi^a \psi^b \right)
\\
&\qquad\qquad
+ \frac{1}{2} \frac{\partial^3 W(\phi)}
{\partial\phi^a \partial\phi^b \partial\phi^c}
\left( F^a \phi^b \phi^c - \phi^a \psi^b \psi^c \right)
+ \hc \biggr]
\eal\eeq
where $W$ is the superpotential of the matter fields.
(We integrate out the auxiliary fields for the gauge multiplet
but not the auxiliary matter fields.)
The Goldstino-dependent terms are
\beq
\eql{matterchione}
\scr{L}^{(1)}_{\rm matter} &=
-2 \partial^\mu \chi \psi^a (\scr{D}_\mu \bar{\phi})_a
+ \hc,
\\
\scr{L}^{(2)}_{\rm matter} &=
-i \partial^\mu \chi \si^\nu \bar{\chi}
(\scr{D}_\mu \bar{\phi})_a (\scr{D}_\nu \phi)^a 
- \partial^\mu \chi \chi F^a (\scr{D}_\mu \bar{\phi})_a
- \partial_\mu \chi \psi^a \scr{D}^\mu (\bar{\chi} \bar{\psi})_a
\nonumber\\
\eql{matterchitwo}
&\qquad
- \frac{g}{2\sqrt{2}} \left[
\partial_\mu \chi \chi \cdot \psi^a \si^\mu (\bar{\la} \bar{\phi})_a
- \partial_\mu \chi \psi^a \cdot \chi \si^\mu (\bar{\la} \bar{\phi})_a
\right] + \hc
\eeq
The gauge kinetic terms give
\beq
\eql{gaugechizero}
\scr{L}^{(0)}_{\rm gauge} &=
-\frac{1}{4} F^{\mu\nu}_A F_{\mu\nu A}
- i \bar{\la}_A \bar{\si}^\mu (\scr{D}_\mu \la)_A
\\
\eql{gaugechione}
\scr{L}^{(1)}_{\rm gauge} &=
-i \sqrt{2} \partial^\mu \chi \si^\nu \bar{\la}_A F_{\mu\nu A}
+ \hc,
\\
\scr{L}^{(2)}_{\rm gauge} &=
\frac{i}{2} \partial_\mu \bar{\chi} \bar{\si}^\nu \chi \cdot
F^{\mu\la}_A (F_{\la\nu A} + i\tilde{F}_{\la\nu A})
- \frac{g_A}{2} \partial^\mu \bar{\chi} \bar{\si}^\nu \chi \cdot
(\bar{\phi} T_A \phi) F_{\mu\nu A}
\nonumber\\
\eql{gaugechitwo}
&\qquad
- \frac{1}{4} \partial_\mu \chi (\scr{D}_\nu \la)_A
\cdot \bar{\chi} \bar{\si}^\nu \si^\mu \bar{\la}_A
- \frac{1}{4} \partial_\mu \chi \si^\nu \bar{\chi}
\cdot (\scr{D}_\nu \la)_A \si^\mu \bar{\la}_A
\nonumber\\
&\qquad
+ \frac{1}{4} \partial_\mu \chi \partial_\nu \chi \cdot
\bar{\la}_A \bar{\si}^\mu \si^\nu \bar{\la}_A
+ \frac{1}{4} \partial_\mu \chi \si^\nu \bar{\la}_A \cdot
\partial_\nu \chi \si^\mu \bar{\la}_A
+ \hc
\eeq
The soft terms give%
\footnote{In an earlier version of this paper, we found a contribution
$\scr{L}^{(2)}_{\rm soft} \propto
\partial^\mu \chi \si^\nu \bar{\chi} F_{\mu\nu} + \hc$
This term is canceled by a similar term from the $D_1$ equation of
motion, giving a result in agreement with \Ref{Purd}. }
\beq
\scr{L}^{(0)}_{\rm soft} &=
-(m^2)^a{}_b \bar{\phi}_a \phi^b
- g_1^2 M_D^2 \bar{\phi} T_1 \phi
\nonumber\\
\eql{softchizero}
&\qquad
+ \left[
\frac{1}{2} B_{ab} \phi^a \phi^b
- \frac{1}{2} \tilde{M}_A \la_A \la_A
+ \frac{1}{6} A_{abc} \phi^a \phi^b \phi^c
+ \hc \right],
\\
\eql{softchitwo}
\scr{L}^{(2)}_{\rm soft} &= 0.
\eeq
Here, $T_1$ is the charge of the $\U1$ factor (see \Eq{SoftTerms}).
We have made extensive use of the $\chi$
equations of motion to simplify these expressions.
The terms linear in $\chi$ are the well-known linear couplings of the
Goldstino to the supercurrent \cite{Fayet}.
% The terms quadratic in $\chi$ are additional
% couplings of the Goldstino to matter.

The fact that the Goldstino couples only through derivatives is an
automatic consequence of the non-linear \rep of \susy we are using.
The couplings above have also been computed by various authors by
taking the low-energy limit of models with linearly realized \susy
\cite{others,CERN}.
In these calculations, the derivative couplings of the Goldstino
arise from cancellations between different terms.
Some results in the literature missed some
of these cancellations, and obtained results in which the Goldstino
is not derivatively coupled.
(We are in agreement with the energy dependencies obtained in
\Ref{CERN}.)
We hope that the derivation presented here is sufficiently
simple and compelling to settle this issue.

% ----------------------------------------------------------------------------
\section{$Z$ Decays into Goldstinos}
% ----------------------------------------------------------------------------
As a simple application of our formalism, we compute the decay width
of the $Z$ into Goldstino pairs.
There are several potential contributions from our effective lagrangian.
The term $\partial^\mu \chi \chi F \scr{D}_\mu \bar{\phi}$ in
\Eq{matterchitwo} gives rise to a coupling of the $Z$ to Goldstino pairs
from terms where $F$ and $\phi$ have \vevs.
However, in the minimal \susc standard model, $\avg{F}$ comes only
from the $H_1 H_2$ term in the superpotential, and
this contribution to the $Z$ coupling vanishes identically.
The term proportional to $D_A$ in \Eq{gaugechitwo} gives a direct
coupling of Goldstino pairs to the $Z$ via
\beq
\de\scr{L} = \frac{1}{2F^2} \partial^\mu \bar{\chi} \bar{\si}^\nu \chi
\avg{D_Y} B_{\mu\nu} + \hc,
\eeq
where
\beq
\avg{D_Y} = \frac{e v^2}{2} \frac{\cos 2\be}{\sin 2\th_W}
\simeq (75\GeV)^2 \left( \frac{\cos 2\be}{0.5} \right).
\eeq
This is small for $\tan\be$ near 1, but we regard this
as a fine-tuned limit.
% This term combines with the induced \FI term discussed above to
% give an effective coupling $M_{DZ}$ defined by
% \beq
% \de\scr{L}\rsub{eff} = \frac{M_{DZ}^2}{2F^2}
% \partial^\mu \chi \si^\nu \bar{\chi} Z_{\mu\nu} + \hc.
% \eeq
% (Note there is no gauge coupling in this definition.)
% If there are no large cancellations, we expect $M_{DZ}^2 \sim \avg{D_Y}$.
This gives a contribution to the invisible width of the $Z$
\beq
\Ga(Z \to \chi\chi) = \frac{\avg{D_Y}^2 M_Z^5}{96 \pi F^4}.
\eeq
We place a limit on $F$ by demanding that this not spoil the
agreement between the invisible $Z$ width as calculated in the
standard model and measured at LEP.
We therefore impose $\Ga(Z \to \chi\chi)< 7.5 \MeV$,
which is 3 times the experimental uncertainty
from the combined LEP average (the theoretical uncertainty is
negligible) \cite{PDG}.
This gives a bound
\beq
\sqrt{F} > (140\GeV) \left( \frac{\cos 2\be}{0.5} \right)^{1/4}.
\eeq

% ----------------------------------------------------------------------------
\section{Conclusions}
% ----------------------------------------------------------------------------
We have computed the low-energy couplings of the longitudinal components
of a light gravitino (the Goldstino) to the minimal \susc standard
model under the assumption that \susy breaking is comminicated from a
hidden sector via soft-breaking terms.
We have shown that these couplings are derivatively coupled to all
orders in the Goldstino field $\chi$, and we have
explicitly worked out the lagrangian to second order in $\chi$.
All couplings of the Goldstino are uniquely determined in terms of
the scale of \susy breaking $\sqrt{F}$.
We used this lagrangian to obtain a weak bound on $\sqrt{F} \gsim 140\GeV$
from the decay of the $Z$ into Goldstinos.
This lagrangian can be used as the starting point for further
phenomenological investigation of the Goldstino in gauge-mediated
and no-scale models.

% ----------------------------------------------------------------------------
\section{Acknowledgments}
% ----------------------------------------------------------------------------
M.A.L. would like to thank R.N. Mohapatra for discussions.

% ----------------------------------------------------------------------------
% References
% ----------------------------------------------------------------------------
%\newpage

\end{document}